\documentclass[pre,amsmath,amssymb,twocolumn]{revtex4}
\usepackage{graphicx}

\begin{document}
\title{Influence of global rotation and Reynolds number on the large-scale features of a turbulent Taylor-Couette flow}

\author{F.~Ravelet}
\affiliation{DynFluid, Arts et M\'etiers-ParisTech, 151 Bld de l'H\^opital, 75013 Paris, France}
\email{florent.ravelet@ensta.org}
\author{R.~Delfos}
\affiliation{Laboratory for Aero and Hydrodynamics, Mekelweg 2, 2628 CD Delft, The Netherlands.}
\author{J.~Westerweel}
\affiliation{Laboratory for Aero and Hydrodynamics, Mekelweg 2, 2628 CD Delft, The Netherlands.}

\date{Submitted to Phys. Fluids: \today}

\begin{abstract}
We experimentally study the turbulent flow between two coaxial and independently rotating cylinders.
We determined the scaling of the torque with Reynolds numbers at various angular velocity ratios (Rotation numbers),
and the behaviour of the wall shear stress when varying the Rotation number at high Reynolds numbers.
We compare the curves with PIV analysis of the mean flow and show the peculiar role of perfect counterrotation
for the emergence of organised large scale structures in the mean part of this very turbulent flow that appear
in a smooth and continuous way: the transition resembles a supercritical bifurcation of the secondary mean flow.
\end{abstract}
\maketitle
\section{Introduction}
\label{sec:intro}

Turbulent shear flows are present in many applied and fundamental problems, ranging from small scales
(such as in the cardiovascular system) to very large scales (such as in meteorology).
One of the several open questions is the emergence of coherent large-scale structures in turbulent flows \cite{holmes}.
Another interesting problem concerns bifurcations, i.e. transitions in large-scale flow patterns
under parametric influence, such as laminar-turbulent flow transition in pipes, or flow pattern change
within the turbulent regime, such as the dynamo instability of a magnetic field in a conducting fluid \cite{monchaux2007},
or multistability of the mean flow in von K\'arm\'an or free-surface Taylor--Couette flows \cite{ravelet2004,mujica2006},
leading to hysteresis or non-trivial dynamics at large scale. 
In flow simulation of homogeneous turbulent shear flow it is observed that there is an important role for what is called
the \emph{background rotation}, which is the rotation of the frame of reference in which the shear flow occurs.
This background rotation can both suppress or enhance the turbulence \cite{tritton92, brethouwer05}.
We will further explicit this in the next section.

A flow geometry that can generate both motions, shear and background rotation, at the same time
is a Taylor-Couette flow, which is the flow produced between differentially rotating coaxial cylinders \cite{couette}.
When only the inner cylinder rotates, the first instability, i.e. deviation from laminar flow with circular streamlines,
takes the form of toroidal (Taylor) vortices. With two independently rotating cylinders, there is a host of interesting
secondary bifurcations, extensively studied at intermediate Reynolds numbers, following the work of Coles \cite{coles1965}
and Andereck \emph{et al.} \cite{andereck1986}. Moreover, it shares strong analogies with Rayleigh-B\'enard convection
\cite{dubrulle2002,eckhardt2007}, which are useful to explain different torque scalings at high Reynolds numbers
\cite{lathrop92,lewis1999}. Finally, for some parameters relevant in astrophysical problems, the basic flow
is linearly stable and can directly transit to turbulence at a sufficiently high Reynolds number \cite{hersant2005}.

\begin{figure}[htbp!]
\begin{center}
\includegraphics[clip,width=.95\columnwidth]{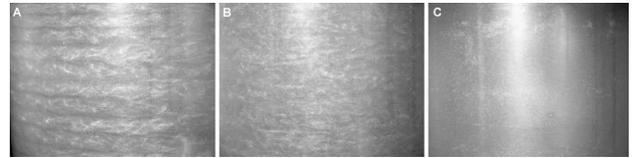}
\caption{Photographs of the flow at $Re=3.6 \times 10^3$. Left, A: only the inner cylinder rotating. Middle, B: counter rotating cylinders.
Right, C: only the outer cylinder rotating. The flow structure is vusualized using microscopic Mica-platelets (Pearlessence).}
\label{fig:photopat}
\end{center}
\end{figure}

The structure of the Taylor-Couette flow while it is in a \emph{turbulent} state, is not so well known and only few measurements
are available \cite{wereley1998}. The flow measurements reported in \cite{wereley1998}
and other torque scaling studies only deal with the case where only the inner cylinder rotates
\cite{lathrop92,lewis1999}. In that precise case, recent direct numerical
simulations suggest that vortex-like structures still exist at high Reynolds number ($Re \gtrsim 10^4$)
\cite{bilson2007,dong2007}, whereas for counter-rotating cylinders, the flows at Reynolds numbers around
$5000$ are identified as \lq\lq featureless states\rq\rq \cite{andereck1986}. The structure of the flow
is exemplified with a flow visualisation in Fig.~\ref{fig:photopat} in our experimental set-up for a flow with
only the inner cylinder rotating, counter-rotating cylinders and only the outer cylinder rotating, respectively.

In the present paper, we extend the study of torques and flow field for independently rotating cylinders
to higher Reynolds numbers (up to $10^5$) and address the question of the transition process between a
\emph{turbulent} flow with Taylor-vortices, and this \lq\lq featureless \rq\rq turbulent flow when varying the global rotation
while maintaining a constant mean shear rate.

In section \ref{sec:tools}, we present the experimental device and the measured quantities. In section \ref{sec:param},
we introduce the specific set of parameters we use to take into account the global rotation through a
\lq\lq Rotation number\rq\rq and the imposed shear through a shear-Reynolds number. We then present torque scalings
and typical velocity profiles in turbulent regimes for three particular Rotation numbers in section \ref{sec:torque}.
We explore the transition between these regimes at high Reynolds number varying the Rotation number in
section \ref{sec:bifurcation}, and discuss the results in section \ref{sec:conclusion}.

\section{Experimental setup and measurement techniques}
\label{sec:tools}

\begin{figure}[htbp!]
\begin{center}
\includegraphics[width=0.85\columnwidth]{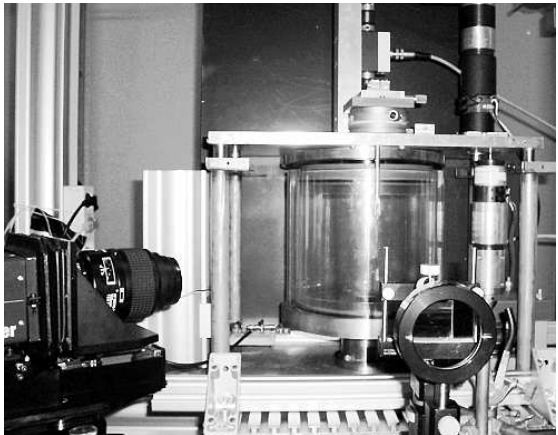}
\includegraphics[width=0.5\columnwidth]{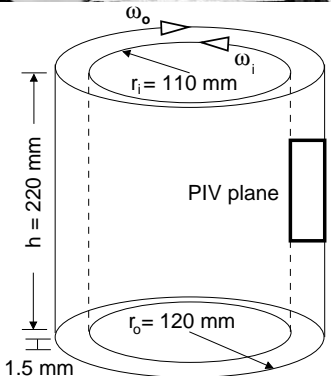}
\caption{Picture and sketch with dimensions of the experimental setup. One can see the rotating torquemeter
(upper part of picture), the calibration grid displacement device (on top of the upper plate), one of the two cameras
(left side) and the light sheet arrangement (right side). The second camera is further to the right.}
\label{fig:manip}
\end{center}
\end{figure}

The flow is generated between two coaxial cylinders (Fig.~\ref{fig:manip}). The inner cylinder has a radius
of $r_i=110 \pm 0.05$~mm, and the outer cylinder of $r_o=120 \pm 0.05$~mm. The gap between the cylinders is thus
$d=r_o-r_i=10$~mm, and the gap ratio is $\eta=r_i/r_o=0.917$. The system is closed at both ends, with top and bottom lids
rotating with the outer cylinder. The length of the inner cylinder is $L=220$~mm (axial aspect ratio is $L/d=22$).
Both cylinders can rotate independently with the use of two DC motors (Maxon, $250$W). The motors are driven by a
home-made regulation device, ensuring a rotation rate up to $10$Hz, with an absolute precision of $\pm 0.02$ Hz and
a good stability. A LabView program is used to control the experiment: the two cylinders are simultaneously
accelerated or decelerated to the desired rotation rates, keeping their ratio constant. This ratio can also be changed
while the cylinders rotate, maintaining a constant differential velocity.

The torque $T$ on the inner cylinder is measured with a co-rotating torquemeter (HBM T20WN, $2$ N.m). The signal is recorded
with a $12$ bits data acquisition board at a sample rate of $2$ kHz for $180$ s. The absolute precision on the torque measurements
is $\pm 0.01$ N.m, and values below $0.05$ N.m are rejected. We also use the encoder on the shaft of the torquemeter to record
the rotation rate of the inner cylinder. Since that matches excellently with the demanded rate of rotation, we assume that
the outer cylinder rotates at the demanded rate as well.

Since the torque meter is mounted in the shaft between driving motor and cylinder, it also records
(besides the intended torque on the wall bounding the gap between the two cylinders) the contribution of mechanical
friction such as in the two bearings, and the fluid friction in the horizontal (K\'arm\'an) gaps between tank bottom and tank top.
While the bearing friction is consiedered to be marginal (and measured so in an empty i.e. air filled system),
the K\'arm\'an-gap contribution is much bigger: during laminar flow,
we calculated and measured this to be of the order of 80$\%$ of the gap torque. Therefore, all measured torques were divided by
a factor 2, and we should consider the scaling of torque with the parameters defined in \S~\ref{sec:param} as more accurate
than the exact numerical values of torque.

A constructionally more difficult, but also more accurate, solution for the torque measurement is to work with three stacked
inner cylinders and only measure the torque on the central section, such as is done in the Maryland Taylor-Couette set-up
\cite{lathrop92}, and (under development) in the Twente Turbulent Taylor-Couette set-up \cite{lohse09}.

We measure the three components of the velocity by stereoscopic PIV \cite{Prasad00} in a plane illuminated by a
double-pulsed Nd:YAG laser. The plane is vertical (Fig.~\ref{fig:manip}), {\em i.e.} normal to the mean flow:
the in-plane components are the radial (u) and axial (v) velocities, while the out-of-plane component is the
azimuthal component (w). It is observed from both sides with an angle of $60^o$ (in air), using two double-frame
CCD-cameras on Scheimpflug mounts. The light-sheet thickness is $0.5$~mm. The tracer particles are 20~$\mu$m
fluorescent (rhodamine B) spheres. The field of view is 11$\times$25~mm$^2$, corresponding to a resolution of
$300 \times 1024$ pixels. Special care has been taken concerning the calibration procedure, on which especially the evaluation
of the plane-normal azimuthal component hevaily relies. As a calibration target we use a thin polyester sheet
with lithographically printed crosses on it, stably attached to a rotating and translating micro-traverse. It is first put into the light sheet
and traversed perpendicularly to it. Typically five calibration images are taken with intervals of $0.5$mm. The raw PIV-images
are processed using Davis$^{\footnotesize{\textcircled{\scriptsize{R}}}}$ 7.2 by Lavision \cite{davis06}. They are first mapped to world coordinates,
then they are filtered with a min-max filter, then PIV processed using a multi-pass algorithm, with a last
interrogation area of $32 \times 32$ pixels with $50\%$ overlap, and normalised using median filtering as post-processing.
Then the three component are reconstructed from the two camera views. The mapping function is a third-order polynomial,
and the interpolations are bilinear. The PIV data acquisition is triggered with the outer cylinder when it rotates,
in order to take the pictures at the same angular position as used during the calibration.

\begin{figure}[tbp!]
\begin{center}
\includegraphics[clip,width=.95\columnwidth]{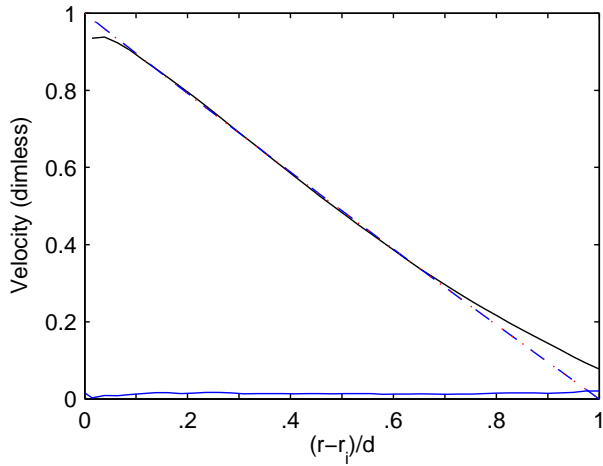}
\caption{Dimensionless azimuthal velocity profile ($w/(r_i \omega_i)$ {\em vs.} $(r-r_i)/d$)
for $Ro=Ro_i$, at $Re_S=90$ (see section \ref{sec:param} for the definition of the parameters).
Solid line: measured mean azimuthal velocity. Dotted line: theoretical profile. Dashed line: fit of the form $w = ar + b/r$.
The radial component $u$ which should be zero is also shown as a thin solid line.}
\label{fig:profilgly}
\end{center}
\end{figure}

To check the reliability of the stereoscopic velocity measurement method, we performed a measurement for a laminar flow
when only the inner cylinder rotates at a Reynolds number as low as $Re_S = 90$, using a $86\%$ glycerol-water mixture. In that case,
the analytical velocity field is known: the radial and axial velocities are zero, and the azimuthal velocity $w$ should be
axisymmetric with no axial dependance, and a radial profile in the form
$w(r) = \Omega r_i \eta  (r_o/r- r/r_o)/{(1-\eta^2)}$ \cite{coles1965}. The results are plotted
in Fig.~\ref{fig:profilgly}. The measured profile (solid line) hardly differs from the theoretical profile (dotted line)
in the bulk of the flow ($0.1 \lesssim (r-r_i)/d \lesssim 0.7$). The discrepancy is however quite strong close to
the outer cylinder ($(r-r_i)/d=1$). The in-plane components which should be zero do not exceed $1\%$ of the inner cylinder
velocity everywhere. In conclusion, the measurements are very satisfying in the bulk. Further improvements to
the technique have been made since this first PIV test, in particular a new outer cylinder of improved roundness,
and the measurements performed in water for turbulent cases are reliable in the range
($0.1 \lesssim (r-r_i)/d \lesssim 0.85$).

\begin{figure*}[htbp!]
\begin{center}
\includegraphics[clip,width=110mm]{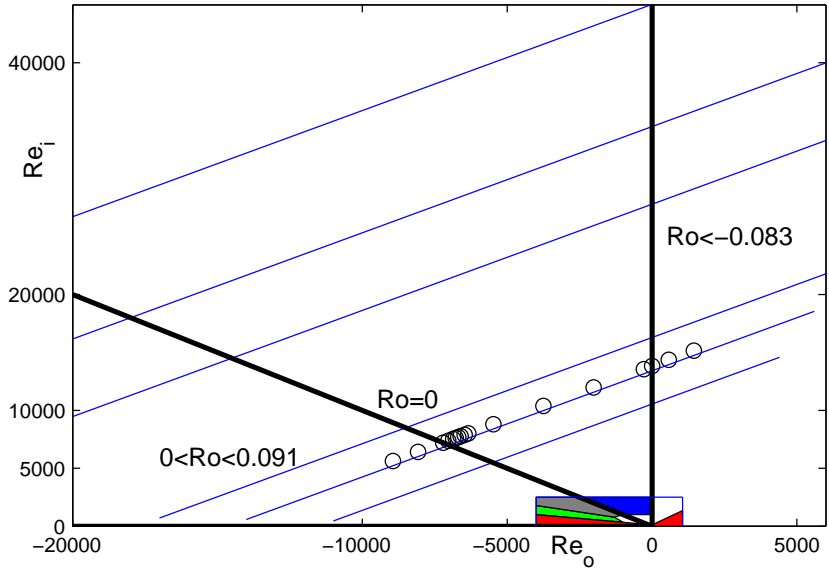}
\caption{Parameter space in $\{Re_o \; ; \; Re_i\}$ coordinates. The vertical axis $Re_o=0$ corresponds to
$Ro=Ro_i=-0.083$, and has been widely studied \cite{lathrop92,lewis1999,bilson2007,dong2007}. The horizontal axis
$Re_i=0$ corresponds to $Ro=Ro_o=0.091$. The line $Re_i=-Re_o$ corresponds to counter-rotation, i.e. $Ro=Ro_c=0$.
The PIV data taken at a constant shear Reynolds number of $Re_S=1.4 \times 10^4$ are plotted with ($\circ$).
Torque data with varying $Ro$ at constant shear for various $Re_S$ ranging from $Re_S=3 \times 10^3$
to $Re_S=4.7 \times 10^4$ are plotted as blue lines.
We also plot the states identified at much lower $Re_S$ by Andereck {\em et al.} \cite{andereck1986} as color patches:
red corresponds to laminar Couette flow, green to \lq\lq spiral turbulence\rq\rq,
grey to \lq\lq featureless turbulence\rq\rq and blue to an \lq\lq unexplored\rq\rq zone.}
\label{fig:chemin2}
\end{center}
\end{figure*}

\section{Parameter space}
\label{sec:param}

The two traditional parameters to describe the flow are the inner (resp. outer) Reynolds numbers,
$Re_{i}=(r_{i} \omega_{i} d /\nu)$ (resp. $Re_{o}=(r_{o} \omega_{o} d /\nu)$), with
the inner (resp. outer) cylinder rotating at rotation rates $\omega_{i}$ (resp. $\omega_{o}$), and
$\nu$ the kinematic viscosity.

We choose to use the set of parameters defined by Dubrulle {\em et al.}~\cite{dubrulle2005}: a shear Reynolds number
$Re_S$ and a \lq\lq Rotation number\rq\rq~$Ro$:

\begin{equation}
\begin{split}
Re_S & = \frac{2|\eta Re_o - Re_i|}{1+\eta} \\
Ro & = (1-\eta) \; \frac{Re_i+Re_o}{\eta Re_o-Re_i}.
\end{split}
\end{equation}

With this choice, $Re_S$ is based on the laminar shear rate $S$: $Re_S=Sd^2/\nu$. For instance with a $20$~Hz velocity
difference in counter-rotation, the shear rate is around $1400$~s$^{-1}$ and $Re_S \simeq 1.4 \times 10^5$ for water
at $20^o$C. A constant shear Reynolds number corresponds to a line of slope $\eta$ in the $\{Re_o \; ; \; Re_i\}$
coordinate system (see Fig.~\ref{fig:chemin2}).

The Rotation number $Ro$ compares the mean rotation to the shear and is the inverse of a Rossby number. Its sign defines
cyclonic ($Ro>0$) or anti-cyclonic ($Ro<0$) flows. The Rotation number is zero in case of perfect counter-rotation
($r_i \omega_i= - r_o \omega_o$). Two other relevant values of the Rotation number are $Ro_i=\eta-1\simeq-0.083$ and
$Ro_o=(1-\eta)/\eta \simeq 0.091$ for respectively inner and outer cylinder rotating alone. Finally, a further choice
that we made in our experiment was the value of $\eta = r_i/r_o$, which we have chosen as relatively close to unity, 
i.e. $\eta = 110/120 \simeq 0.91$, which is considered a narrow-gap, and is the most common in reported experiments,
such as \cite{andereck1986, coles1965, wendt33, racina2006}, though a value as low as 0.128 is described as well \cite{mujica2006}.
A high $\eta$, {\emph i.e.} $(1-\eta) \ll 1$, is special in the sense that for $\eta \rightarrow 1$ a plane Couette flow
with background rotation; at high $\eta$, the flow is linearly unstable for $-1<Ro< Ro_o$ \cite{dubrulle2005, esser1996, tritton92}.

In the present study we experimentally explore regions of the parameter space that, to our knowledge, have not been
reported before. We present in Fig.~\ref{fig:chemin2} the parameter space in $\{Re_o \; ; \; Re_i\}$ coordinates with
a sketch of the flow states identified by Andereck \emph{et al.} \cite{andereck1986}, and the location of the data discussed
in the present paper. One can notice that the present range of Reynolds numbers is far beyond that of Andereck,
and that with the PIV-data we mainly explore the zone between perfect counterrotation and only the inner cylinder rotating.

\section{Study of three particular Rotation numbers}
\label{sec:torque}

In the experiments reported in this section, we maintain the Rotation number at constant values and vary the shear Reynolds number.
We compare three particular Rotation numbers. $Ro_i$, $Ro_c$ and $Ro_o$, corresponding to rotation of the inner cylinder only,
exact counter-rotation and rotation of the outer cylinder rotating only, respectively. In section A we report torque scaling
measurements for a wide range of Reynolds numbers ---from base laminar flow to highly turbulent flows--- and in section B
we present typical velocity profiles in turbulent conditions.

\subsection{Torque scaling measurements}
\begin{figure}[htbp!]
\begin{center}
\includegraphics[clip,width=.95\columnwidth]{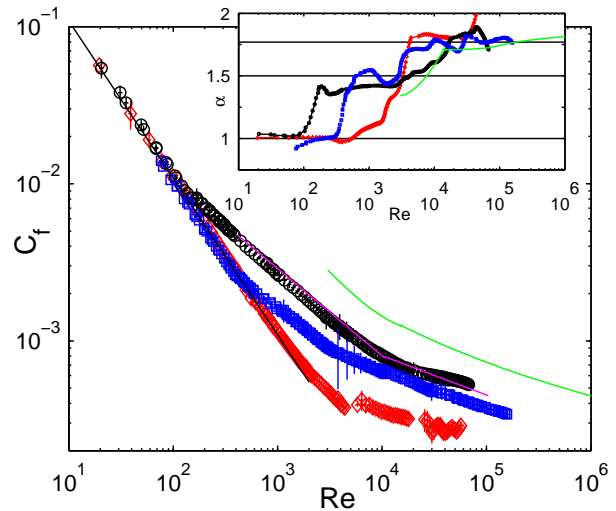}
\caption{Friction factor $c_f$ vs. $Re_S$ for $Ro_i=\eta-1$ (black $\circ$), $Ro_c=0$ (blue $\Box$)
and $Ro_o=(1-\eta)/\eta$ (red $\diamond$). Relative error on $Re_S$: $\pm 5\%$, absolute error on torque: $\pm 0.01$~Nm.
Inset: local exponent $\alpha$ such that $C_f \propto Re_S^{\alpha-2}$, computed as $2+d \log (C_f) / d \log (Re_S)$,
for $Ro_i=\eta-1$ (black), $Ro_c=0$ (blue) and $Ro_o=(1-\eta)/\eta$ (red). Solid green line: Lewis' data,
(Ref.~\cite{lewis1999} eq.~3), for $Ro_i$ and $\eta=0.724$. Solid magenta line: Racina's data (Ref.~\cite{racina2006}, eq.~10).
Solid black line: laminar friction factor $c_f=1/(\eta Re)$.}
\label{fig:scaling}
\end{center}
\end{figure}

We present in Fig.~\ref{fig:scaling} the friction factor $c_f = T/(2\pi \rho r_i^2 L U^2) \propto G/Re^2$,
with $U=Sd$ and $G=T/(\rho L \nu^2)$, as a function of $Re_S$ for the three Rotation numbers. A common definition for
the scaling exponent $\alpha$ of the dimensionless torque is based on $G$: $G \propto Re_S^{\alpha}$. We keep this definition
and present the local exponent $\alpha$ in the inset in Fig.~\ref{fig:scaling}. We compute $\alpha$ by means of a
logarithmic derivative, $\alpha=2+d \log (c_f) / d \log (Re_S)$.

At low $Re$, the three curves collapse on a $Re^{-1}$ curve. This characterizes the laminar regime where
the torque is proportional to the shear rate on which the Reynolds number is based.

For $Ro_i=\eta-1$, one can notice a transition to a different regime at $Re_{ci} \simeq 140$ (the theoretical
threshold is computed as $Re=150$ \cite{esser1996}). This corresponds to the linear instability of the basic flow,
leading in this case to the growth of laminar Taylor vortices. The friction factor is then supposed to scale
as $c_f \propto Re^{-1/2}$ ($\alpha=3/2$), which is the case here (see inset in Fig.~\ref{fig:scaling}).
For exact counter-rotation ($Ro_c=0$), the first instability threshold is $Re_{cc} \simeq 400$.
This is somewhat lower than the theoretical prediction $Re_{cc} = 515$ \cite{esser1996}, which is probably due to
our finite aspect-ratio. Finally, the Taylor-Couette flow with only the outer cylinder rotating ($Ro_o = (1-\eta)/\eta$)
is linearly stable whatever $Re$. We observe the experimental flow to be still laminar up to high $Re$;
then in a rather short range of $Re$-numbers, the flow transits to a turbulent state at $4000 \lesssim Re_{to}
\simeq 5000$.

Further increase of the shear Reynolds number also increases the local exponent (see inset in Fig.~\ref{fig:scaling}).
For $Ro_i=\eta-1$, it gradually rises from $\alpha \simeq 1.5$ at $Re \simeq 200$ to $\alpha \simeq 1.8$ at
$Re \simeq 10^5$. The order of magnitude of these values agree with the results of Lewis {\em et al}
\cite{lewis1999}, though a direct comparison is difficult, owing to the different gap ratios of the experiments.
The local exponent is supposed to approach a value of 2 for increasing gap ratio. Dubrulle \& Hersant
\cite{dubrulle2002} attribute the increase of $\alpha$ to logarithmic corrections, whereas
Eckhardt {\em et al} \cite{eckhardt2007} attribute the increase of $\alpha$ to a balance between a
boundary-layer/hairpin contribution (scaling as $\propto Re^{3/2}$) and a bulk contribution
(scaling as $\propto Re^2$). The case of perfect counter-rotation shows a plateau at
$\alpha \simeq 1.5$ and a sharp increase of the local exponent to $\alpha \simeq 1.75$ at $Re_{tc} \simeq 3200$,
possibly tracing back to a secondary transition. The local exponent
then seems to increase gradually. Finally, for outer cylinder rotating alone ($Ro_o$) the transition is very sharp
and the local exponent is already around $\alpha=1.77$ at $Re \gtrsim 5000$. Note that the dimensional values
of the torque at $Ro_o$ are very small and difficult to measure accurately, and that these may become
smaller than the contributions by the two K\'arm\'an layers (end-effects) that we simply take into account
by dividing by 2 as described in \S~\ref{sec:tools}.
One can finally notice that at the same shear Reynolds number,
for $Re \geq 10^4$ the local exponents for the three rotation numbers are equal within $\pm 0.1$ and that
the torque with the inner cylinder rotating only is greater than the torque in counter-rotation,
the latter being greater than the torque for only the outer cylinder rotating.

\subsection{Velocity profiles at a high shear-Reynolds number}

\begin{figure}[htbp!]
\begin{center}
\includegraphics[clip,height=85mm]{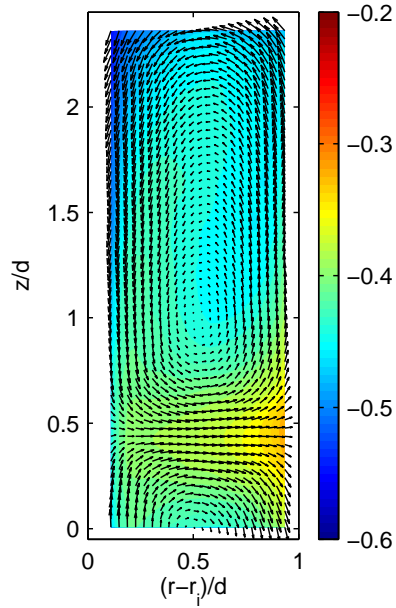}
\caption{Secondary flow for $Ro=Ro_i$ at $Re=1.4 \times 10^4$. Arrows indicate radial and axial velocity,
color indicates azimuthal velocity (normalized to inner wall velocity).}
\label{fig:inneralone}
\end{center}
\end{figure}

\begin{figure}[htbp!]
\begin{center}
\includegraphics[clip,width=.95\columnwidth]{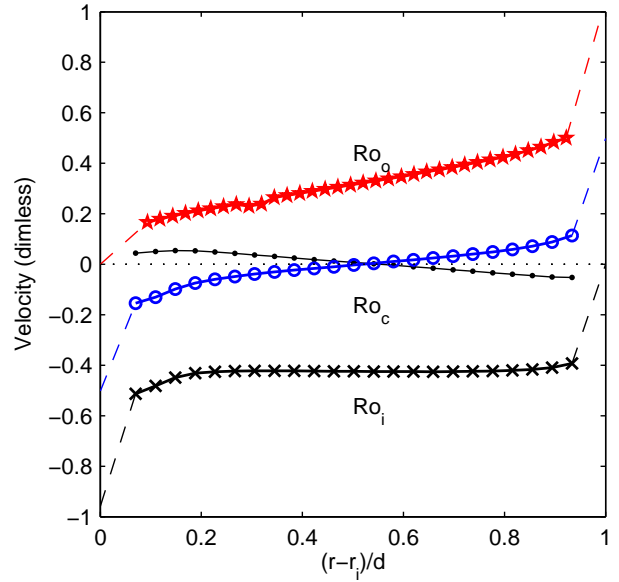}
\caption{Profiles of the mean azimuthal velocity component for three Rotation numbers corresponding to only
the inner cylinder rotating ($\times$, black), perfect counterrotation ($\circ$, blue) and only
the outer cylinder rotating $\star$, red), at $Re=1.4 \times 10^4$.
Thin line ($\cdot$, black)~: axial velocity $v$ (for $Ro=Ro_i$), averaged over half a period.
The velocities are presented in a dimensionless form~: $w/(Sd)$ with $Sd = 2 r_i (\omega_o - \omega_i) /(1+\eta)$.}
\label{fig:profils}
\end{center}
\end{figure}

The presence of vortex-like structures at high shear-Reynolds number ($Re_S \gtrsim 10^4$) in turbulent Taylor-Couette flow
with the inner cylinder rotating alone is confirmed in our experiment through stereoscopic PIV measurements \cite{ravelet2007}.
As shown in Fig.~\ref{fig:inneralone}, the time-averaged flow shows a strong secondary mean flow in the form of counter-rotating
vortices, and their role in advecting angular momentum (as visible in the colouring by the azimuthal velocity)
is clearly visible as well.
The azimuthal velocity profile averaged over both time and axial position, $w$, as shown in Fig.~\ref{fig:profils},
is almost flat, indicating that the transport of angular momentum is due mainly to the time-average coherent structures,
rather than by the correlated fluctuations as in regular shear flow.

We then measured the counter rotating flow, at the same $Re_S$. The measurements are triggered on the outer cylinder position,
and are averaged over 500 images. In the counter rotating case, for this large gap ratio and at this value of the shear-Reynolds number,
the instantaneous velocity field is really desorganised and does not contain obvious structures like Taylor-vortices,
in contrast with other situations \cite{wang2005}. No peaks are present in the time spectra, and there is no axial-dependency
of the time-averaged velocity field. We thus average in the axial direction the different radial profiles; the azimuthal component $w$
is presented in Fig.~\ref{fig:profils} as well. In the bulk it is low, i.e. its magnitude is below $0.1$ between
$0.15 \lesssim (r-r_i)/d \lesssim 0.85$ that is $75\%$ of the gap width. The two other components are zero within $0.002$.

We finally address the outer cylinder rotating alone, again at the same $Re_S$. These measurements are done much in the same way
as the counter-rotating ones, i.e. again the PIV system is triggered by the outer cylinder. As in the counter-rotating flow,
this flow does not show any large scale structures. The gradient in the average azimuthal velocity,
again shown in Fig.~\ref{fig:profils}, is much steeper than in the counter-rotating case, which can be attributed
to the much lower turbulence, as it also manifests itself in the low $c_f$ value for $Ro_o$.

\section{Influence of rotation on the emergence and structure of the turbulent Taylor vortices}
\label{sec:bifurcation}

\begin{figure}[htbp!]
\begin{center}
\includegraphics[clip,width=.95\columnwidth]{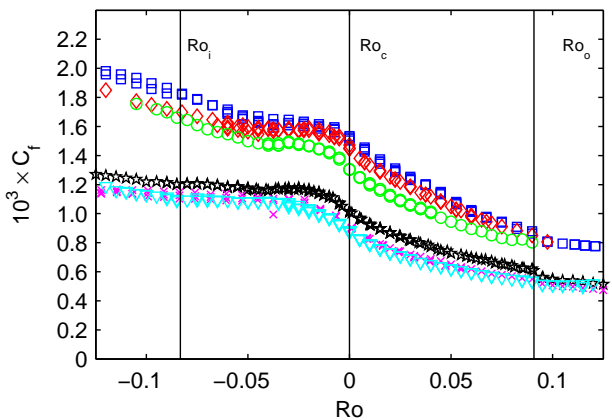}
\caption{The friction factor $c_f$ as a fucntion of $Ro$ at various constant shear Reynolds numbers: (blue) $\Box$ $Re=1.1 \times 10^4$,
(red) $\diamond$ $Re=1.4 \times 10^4$, (green) $\circ$ $Re=1.7 \times 10^4$, (black) $\star$ $Re=2.9 \times 10^4$,
(magenta) $\times$ $Re=3.6 \times 10^4$, (cyan) $\triangledown$ $Re=4.7 \times 10^4$.}
\label{fig:gdero2}
\end{center}
\end{figure}

To characterize the transition between the three flow regimes, we first consider the global torque measurements.
We plot in Fig.~\ref{fig:gdero2} the friction factor or dimensionless torque as a function of Rotation number $Ro$
at six different shear Reynolds numbers, $Re_S$, as indicated in Fig.~\ref{fig:chemin2}. We show three series
centered around $Re_S = 1.4 \times 10^4$, and three around $Re_S = 3.8 \times 10^4$.
As already seen in Fig.~\ref{fig:scaling}, the friction factor reduces with increasing $Re_S$.
More interesting is the behavior of $c_f$ with $Ro$: the torque in counter-rotation ($Ro_c$) is approximately $80\%$
of $c_f(Ro_i)$, and the torque with outer cylinder rotating alone ($Ro_o$) is approximately $50\%$ of $c_f(Ro_i)$.
These values compare well with the few available data, compiled by Dubrulle {\em et al}
\cite{dubrulle2005}. The curve shows a plateau of constant torque especially at the larger $Re_S$
from $Ro = -0.2$, {\emph i.e.} when both cylinders rotate in the same direction with the inner cylinder
rotating faster than the outer cylinder, to $Ro \simeq -0.035$, {\emph i.e.} with a small amount of counter-rotation
with the inner cylinder still rotating faster than the outer cylinder.
The torque then monotonically decreases when increasing the angular speed of the outer cylinder,
with an inflexion point close or equal to $Ro_c$, It is observed that the transition is continuous
and smooth everywhere, and without hysteresis.

\begin{figure}[htbp!]
\begin{center}
\includegraphics[clip,width=.95\columnwidth]{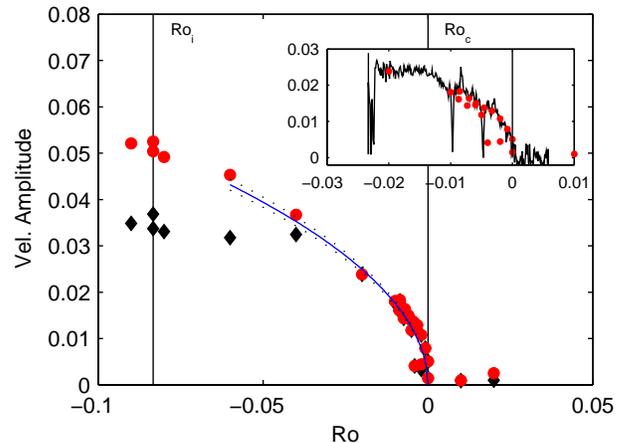}
\caption{Secondary flow amplitude {\em vs.} rotation number ($Ro$) at constant shear rate. Black ($\diamond$):
model with fundamental mode only, and red ($\circ$): complete model with third harmonic. Solid line is a fit
of the form $A=a(-Ro)^{1/2}$. Inset: zoom close to counter-rotation, combined with results from
a continuous transient experiment (see text).}
\label{fig:fitpsi}
\end{center}
\end{figure}

We now address the question of the transition between the different torque regimes by considering the changes observed
in the mean flow. To extract quantitative data from the PIV measurements, we use
the following model for the stream function $\Psi$ of the secondary flow:

\begin{equation}
\begin{split}
\Psi & = \sin\left( \frac{\pi (r\!-\!r_i)}{d}\right) \times \\
     & \times \left[A_1 \, \sin \left(\frac{\pi(z\!-\!z_0)}{\ell} \right)
+A_3\, \sin \left(\frac{3\pi(z\!-\!z_0)}{\ell}]\right) \right]
\end{split}
\end{equation}

with as free parameters $A_1,A_3,\ell$
and $z_0$. This model comprises of a flow that fulfills the kinematic boundary condition at the inner and outer wall,
$r_i, r_i + d$, and in between forms in the axial direction alternating rolls, with a roll height of $\ell$.
In this model, the maximum radial velocity is formed by the two amplitudes
and given by $u_{r,Max} = (\partial \Psi / \partial z)_{Max} = \pi (A_1/\ell +3A_3/\ell)$.
It is implicitly assumed that the flow is developed sufficiently to restore the axisymmetry,
which is checked {\em a posteriori}.
Our fitting model comprises of a sinusoidal (fundamental) mode, and its first symmetric harmonic (third mode), the latter
which appears to considerably improve the matching between the model and the actual average velocity fields,
especially close to $Ro_i$ (see Fig.~\ref{fig:fitpsi}).

\begin{figure}[htbp!]
\begin{center}
\includegraphics[clip,height=85mm]{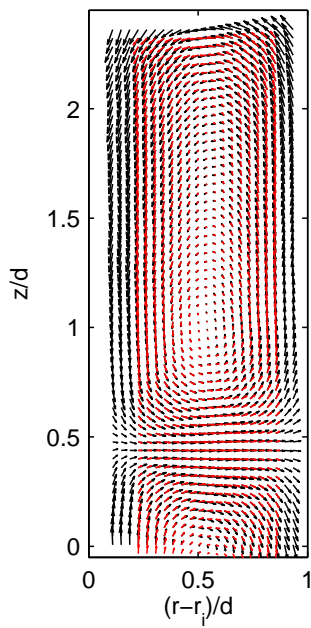}
\caption{Overlay of measured time-average velocity field at $Ro=Ro_i$ (in red), and the best-fit model velocity field (in black).
The large third harmonic makes the radial flow being concentrated in narrow bands, rather than sinusoidally distributed.}
\label{fig:psiexp}
\end{center}
\end{figure}

We first discuss the case $Ro=Ro_i$. A sequence of 4,000 PIV images at a data rate of 3.7~Hz is taken,
and 20 consecutive PIV images, {\em i.e.} approximately 11 cylinder revolutions, are sufficient to obtain a reliable
estimate of the mean flow \cite{dong2007}. It is known that for the first transition the observed flow state can depend on
the initial conditions \cite{coles1965}. When starting the inner cylinder from rest and accelerating it to 2~Hz in 20~s,
the vortices grow very fast, reach a value with a velocity amplitude of $0.08 ms^{-1}$, and then decay to become stabilized
at a value around $0.074 ms^{-1}$ after 400 seconds. Transients are thus also very long in turbulent Taylor-vortex flows.
For slower acceleration, the vortices that appear first are much weaker and have a larger length scale, before reaching
the same final state. The final length scale $\ell$ of the vortices for $Ro_i$ is about 1.2 times the gap width,
consistent with data from Bilson {\em et al.}~\cite{bilson2007}.

In a subsequent measurement we start from $Ro=Ro_i$ and vary the rotation number in small increments,
while maintaining a constant shear rate. We allow the system to spend 20 minutes in each state before acquiring PIV data.
We verify that the fit parameters are stationary, and compute them using the average of the full PIV data set at each $Ro$.
The results are plotted in Fig.~\ref{fig:fitpsi}. Please note that $Ro$ has been varied both with increasing
and decreasing values, to check for a possible hysteresis. All points fall on a single curve; the transition is smooth
and without hysteresis. For $Ro \geq 0$, the fitted modes have zero or negligible amplitudes,
since there are no structures in the time-average field \cite{ravelet2007}. One can notice that as soon as
$Ro < 0$, {\em i.e.} as soon as the inner cylinder wall starts to rotate faster than the outer cylinder wall, vortices begin to grow.
We plot in Fig.~\ref{fig:fitpsi} the velocity amplitude associated with the simple model (single mode $\diamond$),
and with the complete model (modes $1$ and $3$, $\circ$). Close to $Ro=0$, the two models coincide: $A_3 \simeq 0$ and
the mean secondary flow is well described by pure sinusoidal structures. For $Ro\lesssim-0.04$, the vortices start to have
elongated shapes, with large cores and small regions of large radial motions in between adjacent vortices;
the third mode is then necessary to adequately describe the secondary flow. The first mode becomes saturated (i.e. it does not
grow in magnitude) in this region. Finally, we give in Fig.~\ref{fig:fitpsi} a fit of the amplitudes close to
$Ro=0$ of the form: $A=a \, (-Ro)^{1/2}$. The velocity amplitude of the vortex behaves like the square root of the distance
to $Ro=0$, a situation reminiscent to a classical supercritical bifurcation, with $A$ as order parameter,
and $Ro$ as control parameter.

We also performed a \emph{continuous transient} experiment, in which we varied the rotation number quasi-statically
from $Ro$=0.004 to $Ro$=--0.0250 in 3000~s, always keeping the Reynolds number constant at $Re_S = 1.4 \times 10^4$.
The amplitude of the mean secondary flow, computed on sequences
of 20 images, is plotted in the inset of Fig.~\ref{fig:fitpsi}. The curve follows the static experiments
(given by the single points), but some downward peaks can be noticed. We checked that these are not the result of
a fitting error, and indeed correspond to the occasional disappearance of the vortices. Still, the measurements
are done at a fixed position in space. Though the very long time-averaged series leads to well-established stationary 
axisymmetric states, it is possible that the instantaneous whole flow consists of different regions.
Further investigation including time-resolved single-point measurements or flow visualizations need to be done
to verify this possibility.

\section{Conclusion}
\label{sec:conclusion}
The net system rotation as expressed in the Rotation number $Ro$ obviously has strong effects on the torque scaling.
Whereas the local exponent evolves in a smooth way for inner cylinder rotating alone, the counter-rotating case exhibits
two sharp transitions, from $\alpha=1$ to $\alpha \simeq 1.5$ and then to $\alpha \simeq 1.75$.
We also notice that the second transition for counter-rotation $Re_{tc}$ is close to the threshold $Re_{to}$ of turbulence onset
for outer cylinder rotating alone.

The rotation number $Ro$ is thus a secondary control parameter. It is very tempting to use the classical formalism
of bifurcations and instabilities to study the transition between featureless turbulence and turbulent Taylor-vortex
flow at constant $Re_S$, which seems to be supercritical; the threshold for the onset of coherent structures in the mean flow
is $Ro_c$. For anticyclonic flows ($Ro<0$), the transport is dominated by large scale coherent structures,
whereas for cyclonic flows ($Ro>0$), it is dominated by correlated fluctuations reminiscent to those in plane Couette flow.

In a considerable range of $Re_S$, counter-rotation ($Ro_c$) is also close or equal to an inflexion point in the torque curve;
this may be related to the cross-over point, where the role of the correlated fluctuations is taken over by the
large scale vortical structures. The mean azimuthal velocity profiles show there is only a marginal viscous contribution
for $Ro \leq 0$ but of order $10\%$ at $Ro_o$ \cite{ravelet2007}. The role of turbulent vs.~large-scale transport
(of angular momentum) should be further investigated from (existing) numerical or PIV velocity data.
Since torque scaling with $Ro$ as measured at much higher $Re_S$ that that used for PIV does qualitatively not change,
these measurements suggests that the large scale vortices are not only persistent in the flow at higher $Re_S$, but that they
also dominate the dynamics of the flow. An answer to the persistence may be obtained from either more detailed analysis
of instantaneous velocity data or from torque scaling measurements at still higher Reynolds numbers in Taylor-Couette
systems such as are under development \cite{lohse09}.

We are particularly indebted to J.R.~Bodde, C.~Gerritsen and W.~Tax for building up and piloting the experiment.
We have benefited of very fruitful discussions with A.~Chiffaudel, F.~Daviaud, B.~Dubrulle, B.~Eckhardt and D. Lohse.

\bibliographystyle{unsrt}
\bibliography{BiblioDelft3}

\end{document}